\documentclass[aps,prl,
showpacs,superscriptaddress,groupedaddress]{revtex4}  
\usepackage{graphicx}  
\usepackage{dcolumn}   
\usepackage{bm}        
\usepackage{amssymb}   
\usepackage{amsmath}
\usepackage{enumitem}

\begin{document}



\title{Implications of exceptional points for few-photon transport in waveguide quantum electrodynamics}
\author{Shanshan Xu}
 \affiliation{Department of Physics,
Stanford University, Stanford, California 94305}

\author{Shanhui Fan}
\email{shanhui@stanford.edu} \affiliation{Department of Electrical
Engineering, Ginzton Laboratory, Stanford University, Stanford,
California 94305}

\date{\today}

\begin{abstract}
We identify a general connection between the physics of exceptional points in non-Hermitian systems and the few-photon bound states in waveguide quantum electrodynamics (QED) systems. We show that, in waveguide QED systems where the local quantum system exhibits an exceptional point,  the tightest-bound few-photon bound state occurs at the exceptional point. We illustrate this connection with an explicit computation on a waveguide QED system in which a waveguide is coupled to a Jaynes-Cummings system. Our result provides a quantum signature of the exceptional point physics and indicates that the physics of exceptional point can be used to understand and control the photon-photon interaction. 
\end{abstract}

\maketitle


Exceptional points generally occur in the eigensystem of a non-Hermitian matrix. At exceptional points, the matrix becomes
defective and pairs of eigenvalues and eigenstates coalesce.  In optics, the implications of exceptional points have been widely investigated in open systems 
and PT-symmetric systems \cite{bender1998real}, both of which are described by non-Hermitian Hamiltonians \cite{muller2008exceptional,heiss2012physics}. 
As a few examples, nontrivial geometric phase
under cyclic parameter variation around exceptional points has been directly observed in a microwave cavity \cite{dembowski2001experimental}, chaotic optical microcavity \cite{lee2009observation} and Jaynes-Cummings system \cite{choi2010quasieigenstate}.
In laser systems, exceptional points due to non-uniform pumping can strongly affect the above-threshold behavior \cite{liertzer2012pump}. 
However, in all these studies, the signature of the exceptional point is at the classical level.  
There have been few explorations about the implications of exceptional points on the quantum level.  

Separate from the development of non-Hermitian physics, there has been significant recent developments in the field of waveguide quantum electrodynamics (QED). In waveguide QED systems, one couples a local quantum system to a waveguide and study the transport properties of few-photon quantum states in the waveguide
\cite{shen2007strongly,shen2007strongly2,
shi2009lehmann,liao2010correlated,zheng2010waveguide,fan2010input,shi2011two,longo2011few,rephaeli2011few,gonzalez2011entanglement,roy2011two,
kolchin2011nonlinear,zheng2012strongly,rephaeli2012few,ji2012two,zheng2013persistent,liao2013correlated,shi2013two,roy2013two,rephaeli2013dissipation,xu2013analytic,
sanchez2014scattering,li2015scattering,xu2015input,caneva2015quantum,shi2015multiphoton,xu2016fano,xu2017generalized,roy2017colloquium}. 
As a particularly noteworthy development in waveguide QED, it has been noted by Shen and Fan that when two photons are injected into a waveguide QED system, the scattering process can create a two-photon bound state \cite{shen2007strongly}. This prediction has been recently demonstrated experimentally by Firstenberg et al \cite{firstenberg2013attractive}. 
The discovery of two-photon bound state points to the promise of exploring the strongly interacting quantum many-body states of light in waveguide QED systems\cite{chang2008crystallization,liang2018observation}.
  
In this Letter, we identify a general connection between the physics of exceptional point and the physics of few-photon bound state in waveguide QED systems by considering
few-photon transport in a waveguide coupled to a local quantum system. We show that the exceptional point of the effective Hamiltonian of the local quantum system in general 
gives the tightest few-photon bound state in the waveguide, which can be probed experimentally by measuring the few-photon correlation functions. 
We illustrate this connection with an explicit computation on a waveguide QED system in which the local quantum system is a Jaynes-Cummings system.
Our work points to a general connection between the non-Hermitian physics and the waveguide QED that has not been explored previously. 
The results indicate that the physics of exceptional point can be used to control photon-photon interaction, and therefore such physics may prove useful in the quest to create many-body quantum photon states in waveguide QED systems. 

We start with a general Hamiltonian of the waveguide QED system 
\begin{equation}\label{fullH}
H=\int dk\, k\,c^{\dag}_kc_k+\sqrt{\frac{\kappa}{2\pi}} \int dk\left(c_k^{\dag}a+a^{\dag}c_k\right)+H_{\text{loc}}\,,
\end{equation}
where $c_k\, (c_k^{\dag})$ are annihilation (creation) operators of photon states in the waveguide that satisfies the
standard commutation relations $\left[c_k, c_{k'}^{\dag}\right]=\delta(k-k')$.  Here for simplicity we consider a waveguide consisting of only a
single mode in the sense of Ref. \cite{shen2007strongly2}. The discussion here, however, can be straightforwardly generalized to waveguides supporting multiple modes.
We consider only a narrow range of frequencies in which the waveguide dispersion relation can be linearized, and the group velocity of the waveguide is taken
to be $1$.  $H_{\text{loc}}$ is the Hamiltonian of the local quantum system. $a (a^{\dag})$ is one of the local system's operators that couples to the waveguide with coupling constant $\kappa$. In practice, $a$ can either be a bosonic operator describing a cavity mode or a spin operator for atom-waveguide interaction. 
When coupled to a waveguide, the local quantum system becomes an open system whose dynamics can be described a non-Hermitian effective Hamiltonian. 
Specifically, for the waveguide-cavity coupling as described in Hamiltonian (\ref{fullH}), it has been shown that the effective Hamiltonian takes the form of \cite{shi2011two,xu2015input}
\begin{eqnarray}
H_{\text{eff}}=H_{\text{loc}}-i\,\frac{\kappa}{2}\,a^{\dag}a\,. \label{effH1}
\end{eqnarray}
Here, the imaginary part of the effective Hamiltonian arises since the waveguide degrees of freedom that couples to the local system forms a continuum. 
We further assume that there exists a conserved excitation number operator $N_{\text{loc}}$ for the local system, satisfying $[N_{\text{loc}}, H_{\text{eff}}] = 0$. As a result,
$H_{\text{eff}}$ in  (\ref{effH1}) has eigenstates 
\begin{eqnarray}\label{eigen1}
H_{\text{eff}}\,|n,\lambda\rangle = E_{n, \lambda}|n,\lambda\rangle\,,\,\,\,\,\,\,\, N_{\text{loc}}\,|n,\lambda\rangle = n\,|n,\lambda\rangle\,,
\end{eqnarray}
where $n\in \mathbb{Z}^+$ is the total excitation number and $\lambda$ denotes different eigenstates with the same excitation number. 
Here, we focus on the local quantum systems having a two-dimensional single-excitation subspace spanned by eigenstates $|1,+\rangle$ and $|1,-\rangle$. The systems of this kind include, for example, 
a Jaynes-Cummings system \cite{shi2011two,rephaeli2012few}, a three-level $V$-shape atom \cite{witthaut2010photon} or a pair of colocated two-level atoms \cite{rephaeli2011few}. 
The eigenvalues $E_{1,+}$ and $E_{1,-}$ are in general in pairs.
However, at exceptional point, the effective Hamiltonian is defective and the pair of eigenvalues $E_{1,\pm}$  coalesce. 
The existence of such an exceptional point in open quantum systems has been noted previously \cite{dembowski2001experimental,lee2009observation,choi2010quasieigenstate}. 

We consider the implication of the existence of the exceptional point for the few photon transport properties for the full Hamiltonian (\ref{fullH}).
For simplicity, we first analyze the case of two-photon transport and then extend our discussion to the $N$-photon case. 
If we inject two photons in the waveguide, these photons will propagate along the waveguide, interact with the local quantum system and output a two-photon bound state. Such a bound state
was first discovered by Shen and Fan in the waveguide QED system where the local quantum system is a single two-level atom \cite{shen2007strongly}. 
Intuitively, the two-photon bound state occurs due to the photon-photon interaction as induced by the two-level atom. Later, it was found that the two-photon
bound state exists in many other waveguide QED systems including the cases where the local quantum system is Kerr-nonlinear cavity \cite{liao2010correlated}, optomechanical cavity \cite{liao2013correlated}, three-level atom \cite{roy2011two,xu2017generalized} and the Jaynes-Cummings system \cite{shi2011two,rephaeli2012few},
as long as there exists nonlinearity in the Hamiltonian of the local quantum system that couples to the waveguide.  To compute the two-photon bound state in these systems,
one can first evaluate the two-photon scattering matrix (S matrix) 
\begin{eqnarray}\label{twoSD}
S_{p_1p_2k_1k_2} = \langle p_1,p_2|\hat{S}|k_1,k_2\rangle\,\nonumber
\end{eqnarray}
that relates the incident photons with frequencies  $k_1, k_2$ to the outgoing photons with frequencies $p_1, p_2$. 
The two-photon S matrix can in general be decomposed into the form 
\begin{eqnarray}
S_{p_1p_2k_1k_2}=S^0_{p_1p_2k_1k_2}+S^C_{p_1p_2k_1k_2}\,.\label{S}
\end{eqnarray}
The first term $S^0_{p_1p_2k_1k_2}$ is the non-interacting part. In the case where the local quantum system has a unique ground state,
$S^0_{p_1p_2k_1k_2} = t_{k_1}t_{k_2}\left[\delta(p_1-k_1)\delta(p_2-k_2)+\delta(p_1-k_2)\delta(p_2-k_1)\right]$,
which describes the process in which each photon transports independently with transmission amplitudes $t_{k_1}$ and $t_{k_2}$. $S^C_{p_1p_2k_1k_2}$ is the interacting part
that describes the interaction between the two photons \cite{xu2013analytic}. For the input of two photons with frequencies $k_1,k_2$, the wavefunction of the output bound state
 is \cite{zheng2010waveguide,xu2016fano}
\begin{equation}\label{boundstate}
B(x_1,x_2)=\frac{1}{4\sqrt{2}\pi}\int dp_1dp_2\, S^C_{p_1p_2k_1k_2}\left(e^{ip_1x_1}e^{ip_2x_2}+e^{ip_1x_2}e^{ip_2x_1}\right)\,.
\end{equation} 
Furthermore, it has been argued that $S^C_{p_1p_2k_1k_2}$ has the analytic structure \cite{xu2013analytic,xu2017generalized}
\begin{equation}\label{connected1}
S^C_{p_1p_2k_1k_2} = \frac{{\cal{A}}(p_1,p_2,k_1,k_2)\delta(p_1+p_2-k_1-k_2)}{\prod_{l=1}^2\prod_{\lambda=\pm}(p_l-E_{1,\lambda})(k_l-E_{1,\lambda})\prod_{\rho}(k_1+k_2-E_{2,\rho})}\,,
\end{equation}
where ${\cal{A}}(p_1,p_2,k_1,k_2)$ is an analytic function on photon frequencies. 
(As has been shown in Ref. \cite{xu2017generalized}, this form is true independent of whether the local quantum system has one or multiple ground states.)
Note that the eigenvalues of the effective Hamiltonian, $E_{1,\pm}$ and $E_{2,\rho}$ as shown in (\ref{eigen1}), correspond to the single- and two-photon excitation poles in the interacting part of the two-photon S matrix (\ref{connected1}) at the lower half of the complex energy plane \cite{xu2013analytic,xu2017generalized}. 
Since the eigenvalues $E_{1,\pm}$ coalesce at the exceptional points of the effective Hamiltonian, we expect that the exceptional point should play a role in the two-photon transport and two-photon bound state as well. 

To explicitly connect the exceptional points in the non-Hermitian physics of the effective Hamiltonian (\ref{effH1}) to the two-photon bound state in the waveguide QED physics, 
\begin{figure}[h]
\centering
\includegraphics[width=0.8\textwidth] {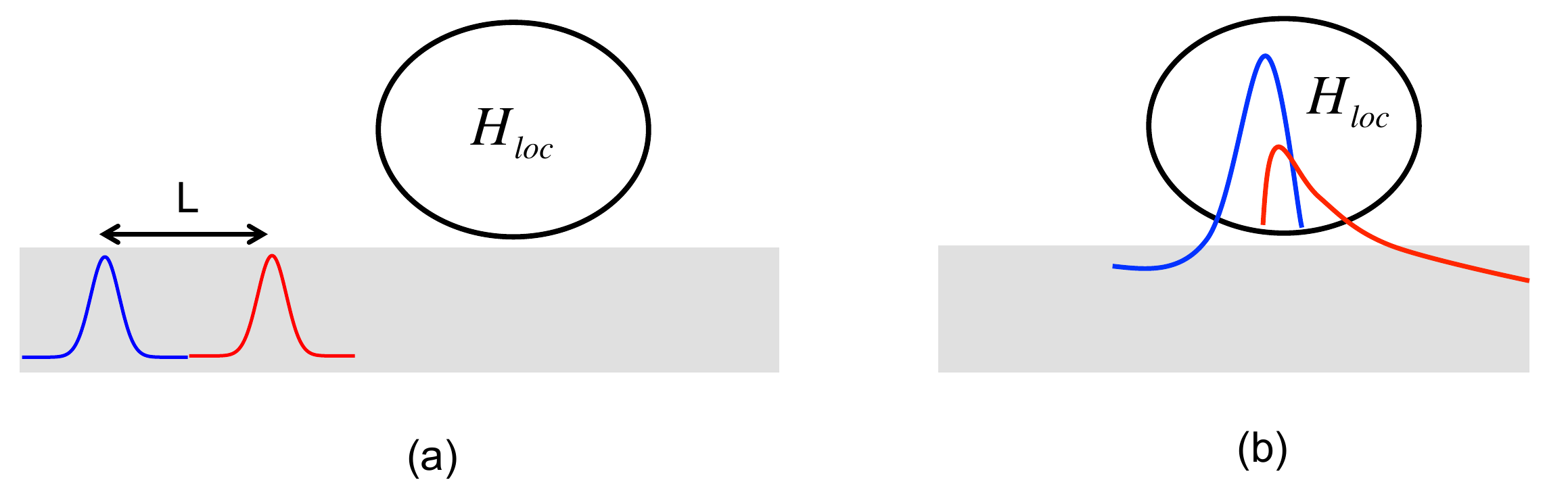} 
\caption{(a) Two single-photon pulses with a separation $L$ scatter against the local quantum system. (b) After the first single-photon pulse passes, the excitation in the local quantum system decays exponentially. The strength of the interaction depends on the amplitudes of such excitation when the second single-photon pulse arrives. } \label{jcfig}
\end{figure}
we consider the scattering process of two single-photon pulses against the local quantum system. 
The two single-photon pulses have the same group velocity $v_g$ but with a separation $L$ as shown in Fig. \ref{jcfig} (a). 
The first pulse excites the local quantum system,  and then
 the amplitude of excitation inside the local system decays into the waveguide in the form of $A_+\, e^{-iE_{1,+}t}+A_-\, e^{-iE_{1,-}t}$ as controlled by the  
 single excitation poles $E_{1,\pm}$ in the single photon transmission amplitude.
The interaction between the photons can occur only if there remains excitation in
the local system at the moment when the second photon pulse arrives. At that moment, the remaining excitation inside the local system 
should be  $A_+\, e^{-iE_{1,+}\tau}+A_-\, e^{-iE_{1,-}\tau}$ where $\tau \equiv L / v_g$
is the time decay between two photons (Fig. \ref{jcfig} (b)). As a result, we
expected that the outcome of such an interaction
should decay as a function of the separation $L$ between the two pulses in the form of $A_+\, e^{-iE_{1,+}L}+A_-\, e^{-iE_{1,-}L}$ when $L$ is large. 
At the exceptional point where $E_{1+}=E_{1-}$, such form exhibits the feature of the critical damping, that is, the interaction decays quickest to zero. 
Note that the two-photon bound state is from such photon-photon interaction, we thus expect intuitively that the critical damping at the exceptional point leads to the tightest two-photon bound state.  
Indeed, we can verify such an intuition by computing the wavefunction of the two-photon bound state explicitly from (\ref{boundstate}) and (\ref{connected1}). For $E_{1,+}\neq E_{1,-}$, we have
\begin{equation}\label{decay1}
B(x_1,x_2) \propto \sum_{\lambda=\pm} {{\cal{A}}_{\lambda}e^{-iE_{1,\lambda}|x_1-x_2|}}\,,
\end{equation} 
while at the exceptional point $E_{1+}=E_{1-}$,
\begin{equation}\label{cdecay}
B(x_1,x_2) \propto \left[1-i\left(\frac{k_1+k_2}{2}-E_{1,+}\right)\left|x_1-x_2\right|\right]{e^{-iE_{1,+}|x_1-x_2|}}\,.
\end{equation} 
Eq. (\ref{cdecay}) is the form of critical damping as a function of photons' separation $|x_1-x_2|$, suggesting that two-photon bound state is tightest compared 
when the system exhibits an exceptional point.

The above relation between the exceptional point and the tightest two-photon bound state also applies to cases where there are more than two photons. In general, as proved in the supplement
material, the wavefunction of the $N$-photon bound state has a pairwise decay form in terms of photons' separations
\begin{equation}
B(x_1, \cdots, x_N) \propto \sum_{Q} \prod_{i=1}^{N-1} D_i(x_{Q(i)}-x_{Q(i+1)})\theta(x_{Q(i)}-x_{Q(i+1)})\,,
\end{equation} 
where $Q$ represents all the permutations of indices $\{1, \cdots, N\}$. $ D_i(x_{Q(i)}-x_{Q(i+1)})$ is a linear combination of 
decay terms controlled by excitation poles $E_{n,\lambda}$ in (\ref{eigen1}) for $n=1$ up to $N$. Such a pairwise decay form has been explicitly 
computed in special cases of multiple-level atoms \cite{zheng2010waveguide,zheng2012strongly}. In waveguide QED systems, one typically has $\text{Im} E_{n,\rho} < \text{Im} E_{1,\pm} <0$ for $n>1$.
The slowest decay is from the single excitation poles and asymptotically we have 
\begin{equation}\label{Nasy}
D_i(x_{Q(i)}-x_{Q(i+1)}) \propto \sum_{\lambda=\pm} {{\cal{A}}_{i,\lambda}e^{-iE_{1,\lambda}|x_{Q(i)}-x_{Q(i+1)}|}}\,.
\end{equation}
The coefficient ${\cal{A}}_{i,\lambda}$ in (\ref{Nasy}) can be calculated from the part of the connected $N$-photon S matrix that only describes the single excitation processes.
As proved in the supplementary material, such S matrix is the product of a single off-shell two-photon S matrix and a series of single photon S matrix.
As a result, the $N$-photon bound state should also be tightest at the exceptional point $E_{1,+}=E_{1,-}$. Our result here suggests that the presence of the exceptional point 
manifests in strongly correlated many-body state of photons.  

\begin{figure}[h]
\centering
\includegraphics[width=0.9\textwidth]{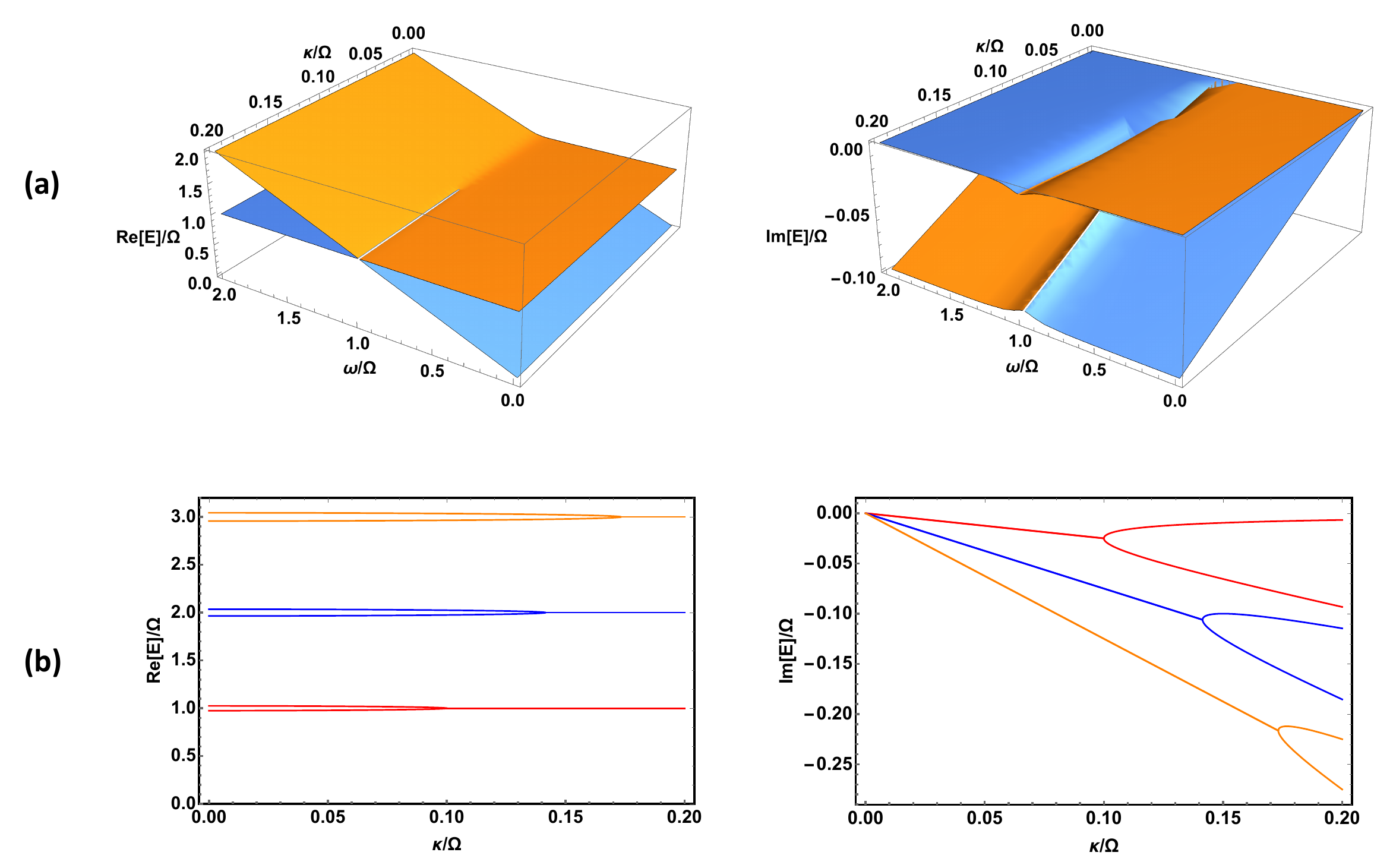}
\caption{(a) The $\omega$ and $\kappa$ dependence of the real of imaginary part of eigenvalues $E_{1,\pm}$ in Eq. (\ref{eigen}) when $g=0.025 \,\Omega$.
The exceptional point is located on the curve where the eigenvalue surfaces intersect $\omega=\Omega, \kappa=0.1\Omega$.
(b) The real and imaginary parts of eigenvalues in the subspaces of excitation numbers $n=$ 1 (red), 2 (blue) and 3 (orange) as a function of $\kappa$. $\omega=\Omega$ and $g=0.025\,\Omega$.}\label{fig2d}
\end{figure}

To support the general argument, we perform an explicit computation for the case where the local quantum system is described by the Jaynes-Cummings Hamiltonian
\begin{equation}\label{jc}
H_{\text{loc}}=\omega \,a^{\dag}a+\Omega\, \sigma_+\sigma_-+g\left[a^{\dag}\sigma_-+\sigma_+a\right]\,,\nonumber
\end{equation}
where $a\, (a^{\dag})$ is the annihilation (creation) operator
of the cavity mode with frequency $\omega$. $\sigma_{\pm}$ are operators of the two-level atom defined by the Pauli matrices $\frac{1}{2}\left(\sigma_x\pm i\sigma_y\right)$. 
$\Omega$ is the atomic transition frequency and  $g$ is the atom-cavity coupling rate.
In this case, the non-Hermitian effective Hamiltonian (\ref{effH1}) and the excitation number operator take the forms of
 \begin{eqnarray}
H_{\text{eff}}&=&\left(\omega-i\,\frac{\kappa}{2}\right) \,a^{\dag}a+\Omega\, \sigma_+\sigma_-+g\left[a^{\dag}\sigma_-+\sigma_+a\right]\,,\label{device}\\
N_{\text{loc}}&=&a^{\dag}a+\sigma_+\sigma_-\,.\nonumber
\end{eqnarray} 
As a result, the eigenvalues in (\ref{eigen1}) are
\begin{equation}\label{eigen}
E_{n,\pm}=\frac{(2 n-1)\left(\omega-i\frac{\kappa}{2}\right)+\Omega}{2}\pm\sqrt{\left(\frac{\omega-i\frac{\kappa}{2}-\Omega}{2}\right)^2+n\,g^2}\,,
\end{equation} 
 In general, for each $n$, there is a pair of eigenvalues.  However, at exceptional point, where $\omega=\Omega$ and $\kappa=4 \sqrt{n}\,g$, 
the Hamiltonian (\ref{device}) is defective and the pair
of eigenvalues (\ref{eigen})  coalesce.  As shown in Fig.\ref{fig2d} (a), in the vicinity of exceptional points, the eigenvalue surfaces form intersecting Riemann sheets in terms of parameters $\omega$ and $\kappa$, leading to a nontrivial geometric phase under cyclic parameter variation in the parameter space \cite{heiss1999phases}.
If we fix $\omega=\Omega$ and vary $\kappa$, as shown in Fig.\ref{fig2d} (b), the eigenvalues (\ref{eigen}) 
coalesce and  exhibit slope discontinuity at the exceptional points $\kappa = 4\sqrt{n}\,g$ for each $n\in \mathbb{Z}^+$. 
Also, we see that indeed $\text{Im} E_{n,\pm} < \text{Im} E_{1,\pm} < 0$ for $n > 1$, confirming a condition required above for the general argument. 
All these behaviors related to the existence of exceptional points have been observed experimentally in an open Jaynes-Cummings system \cite{choi2010quasieigenstate}. 

\begin{figure}[h]
\centering
\includegraphics[width=1.0\textwidth]{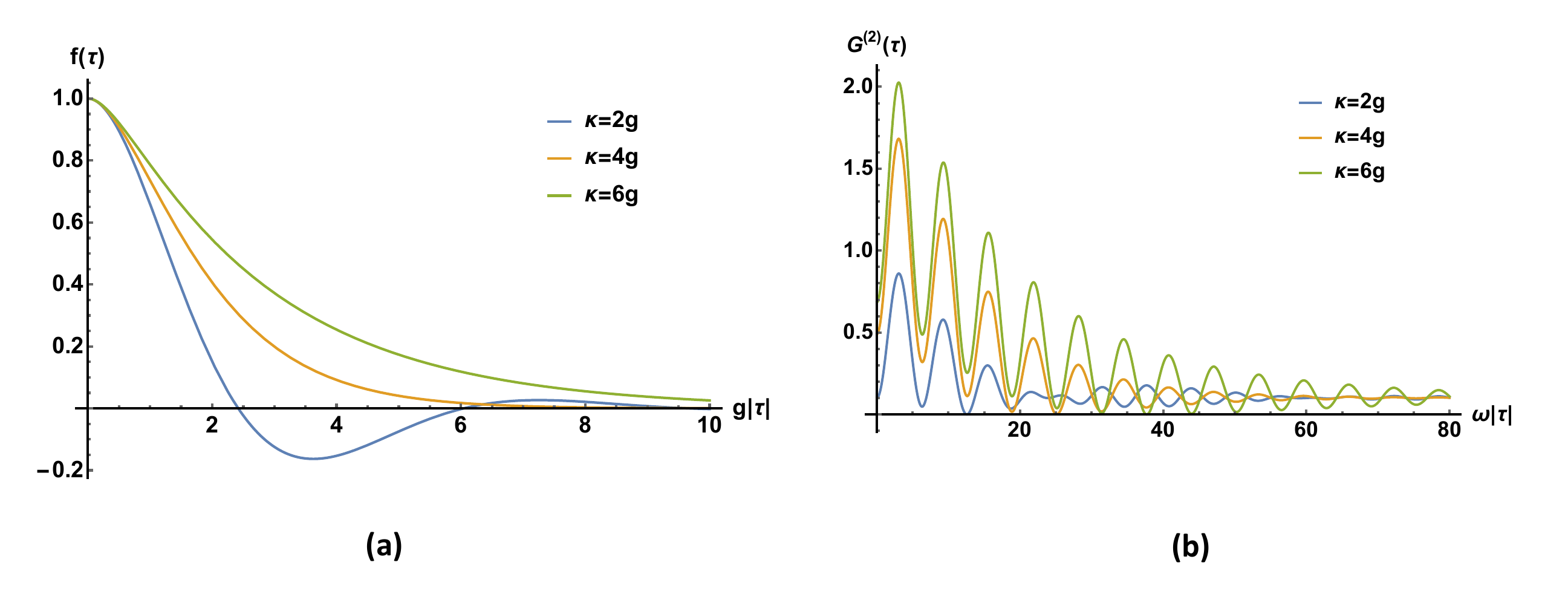}
\caption{(a) The profile  of the two-photon bound state as a function of the separation between two photons in different regions separated by the exceptional point.
(b) The two-photon correlation function $G^{(2)}(\tau)$ in different regions separated by the exceptional point. $g$ is fixed to be $0.1\omega$}\label{ftau}
\end{figure}

The exceptional point behavior as indicated above manifests in the few photon transport properties. 
As calculated in Ref. \cite{shi2011two,rephaeli2012few},
The single photon transmission coefficient has the form 
\begin{equation}
t_k=\frac{\left(k-\omega-i\frac{\kappa}{2}\right)\left(k-\Omega\right)-g^2}{\left(k-\omega+i\frac{\kappa}{2}\right)\left(k-\Omega\right)-g^2}\,,\label{singleS}
\end{equation} 
and the interacting part of the two-photon S-matrix is
\begin{equation}
S^C_{p_1p_2k_1k_2}=\frac{{\kappa}g^2F(k_1, k_2)\,\delta(p_1+p_2-k_1-k_2)}{(p_1-E_{1,+})(p_1-E_{1,-})(p_2-E_{1,+})(p_2-E_{1,-})}\,,\label{connected}
\end{equation}
with
\begin{equation}
F(k_1,k_2)\equiv\frac{i\sqrt{\kappa}g}{\pi}\frac{2g\left[s_{k_1}^{(c)}+s_{k_2}^{(c)}\right]+(k_1+k_2-2\omega+i\kappa)\left[s_{k_1}^{(a)}+s_{k_2}^{(a)}\right]}{(k_1+k_2-E_{2,+})(k_1+k_2-E_{2,-})}\,,\nonumber
\end{equation}
and 
$s_{k}^{(c)}\equiv\frac{\sqrt{\kappa}(k-\Omega)}{\left(k-\omega+i\frac{\kappa}{2}\right)\left(k-\Omega\right)-g^2}$,
$s_{k}^{(a)}\equiv\frac{\sqrt{\kappa}g}{\left(k-\omega+i\frac{\kappa}{2}\right)\left(k-\Omega\right)-g^2}$.
In (\ref{connected}), we have $E_{1,+}=E_{1,-}$ at the exceptional point $\omega=\Omega,\,\kappa=4g$. As a result, $p_1$ and $p_2$ have a double pole instead of two single poles, which
results in the critical damping as discussed above. In a special case of resonant scattering where two input photons have the same frequency $k_1=k_2=\omega=\Omega$,
we can compute the wavefunction of the two-photon bound state explicitly from (\ref{boundstate}) and (\ref{connected}) as
\begin{equation}\label{bs}
B(x_c,\tau)=-\frac{4\kappa^2}{\sqrt{2}\pi(\kappa^2+4g^2)}e^{2i\omega x_c} f(\tau)\,,
\end{equation}
where $x_c$ is the coordinate of the  two-photon center-of-mass $x_c\equiv \frac{x_1+x}{2}$ and $\tau$ is the spatial separation $\tau\equiv x_1-x_2$. $f(\tau)$ has different forms depending on the values of $\kappa$ and $g$:
\begin{equation}\label{ftau1}
f(\tau)=\begin{cases}
\left(\cos\sqrt{g^2-\left(\frac{\kappa}{4}\right)^2}|\tau|+\frac{\kappa\sin\sqrt{g^2-\left(\frac{\kappa}{4}\right)^2}|\tau| }{\sqrt{(4g)^2-\kappa^2}}\right)e^{-\frac{\kappa}{4}|\tau|}  & \kappa<4g  \\
\left(1+g\,|\tau| \right)e^{-g|\tau|} \ &\kappa=4g \\
\left(\cosh\sqrt{\left(\frac{\kappa}{4}\right)^2-g^2}|\tau|+\frac{\kappa\sinh\sqrt{\left(\frac{\kappa}{4}\right)^2-g^2}|\tau| }{\sqrt{\kappa^2-(4g)^2}}\right)e^{-\frac{\kappa}{4}|\tau|}  & \kappa>4g  \\
\end{cases}\,,
\end{equation}
which is a special form of (\ref{decay1}) and (\ref{cdecay}).
The behavior of $f(\tau)$ is shown in Fig. \ref{ftau} (a). At $\kappa = 4g$, $f(\tau)$ is critically damped and has the smallest spatial extent. At $\kappa < 4g$, $f(\tau)$ is underdamped, it oscillates as a function of $\tau$. At $\kappa > 4g$, $f(\tau)$ is over damped, it decays to zero exponentially as $\tau$ increases. For both $\kappa > 4g$ and $\kappa < 4g$, the spatial extent of $f(\tau)$ is larger as compared to the critically damped case with $\kappa = 4g$. The result here illustrates that the exceptional point in waveguide QED systems has a quantum signature in the properties of the two-photon bound state of the system. 

The properties of the two-photon bound state can be probed experimentally by measuring the two-photon correlation function which 
is defined as
\begin{eqnarray}\label{G2}
G^{(2)}(\tau)=\langle \text{out} | c^{\dag}(x) c^{\dag}(x+\tau)c(x+\tau)c(x)|\text{out}\rangle\,.
\end{eqnarray}
Here $|\text{out}\rangle\equiv \hat{S} |k_1,k_2\rangle$ is the out state after scattering when the input state consists of two photons with frequencies $k_1,k_2$. $c(x)$ and $c^{\dag}(x)$
are the annihilation and creation operators of the waveguide photons in the coordinate representation, satisfying the commutation relation $\left[c(x),c^{\dag}(y)\right]=\delta(x-y)$.
Again, we consider the case of resonant scattering where $k_1=k_2=\omega=\Omega$. The two-photon correlation function can be computed explicitly from (\ref{G2}) and (\ref{S})-(\ref{connected}) as
\begin{equation}\label{G2s}
G^{(2)}(\tau)=\left|\frac{1}{\pi}e^{i\omega\tau}-\frac{4\kappa^2}{\pi\left(\kappa^2+4g^2\right)}f(\tau)\right|^2\,,
\end{equation}
where the first term in the absolute value arises due to the non-interacting part of S matrix (\ref{S}).
$f(\tau)$ appears because of the two-photon bound state contained in the state $|\text{out}\rangle$.   
Due to the contribution of the non-interacting part of the S matrix, $G^{(2)}(\tau)$ in general oscillates as a function of $\tau$ and approaches $1/\pi$ when $\tau\rightarrow\infty$, as shown in Fig. \ref{ftau} (b) where we plot the function $G^{(2)}(\tau)$ for different values of $\kappa/g$. 
We note that at $\kappa = 4g$, where the effective Hamiltonian supports an exceptional point in the single-excitation subspace, the approach of $G^{(2)}(\tau)$ to the asymptotic value of $1/\pi$ is the quickest. Therefore, there is an experimental signature of the exceptional point in the two-photon correlation function. 

For the $N$-photon scattering with resonate frequency $k_1=k_2\cdots=k_N=\omega=
\Omega$, as calculated in the supplementary material, the output $N$-photon bound state has the asymptotic form of 
\begin{eqnarray}
B(x_1, \cdots, x_N) \propto \sum_{Q}f(x_{Q(j)}-x_{Q(j+1)})\theta(x_{Q(1)}-x_{Q(2)})\prod_{j=2}^{N-1}g(x_{Q(j)}-x_{Q(j+1)})\theta(x_{Q(j)}-x_{Q(j+1)})
\end{eqnarray}
with $f(\tau)$ defined in (\ref{ftau1}) and $g(\tau)$ defined as 
\begin{equation}\label{gtau}
g(\tau)\equiv \begin{cases}
\frac{\sin\sqrt{g^2-\left(\frac{\kappa}{4}\right)^2}|\tau| }{\sqrt{(g)^2-\left(\frac{\kappa}{4}\right)^2}}e^{-\frac{\kappa}{4}|\tau|}   & \kappa<4g  \\
|\tau| e^{-g|\tau|} \ &\kappa=4g \\
\frac{\sinh\sqrt{\left(\frac{\kappa}{4}-g^2\right)^2}|\tau| }{\sqrt{\left(\frac{\kappa}{4}\right)^2-g^2}}e^{-\frac{\kappa}{4}|\tau|} & \kappa>4g  \\
\end{cases}\,.
\end{equation}
The $N$-photon bound state also exhibits critical damping and thus is tightest at the exceptional point $\kappa=4g$.

In summary, we consider the few-photon transport in a waveguide coupled to a local quantum system. We show that the exceptional point in the open quantum local system has a direct signature in the few-photon bound state of in the waveguide. The tightest-bound few-photon bound state in this system occur at the exceptional point. This connection between the tightest-bound photon bound state, and the exceptional point of the open system, is a general one, since it arises from the critical damping property that occurs at the exceptional point. Our work points to a connection between the non-Hermitian physics and the waveguide QED that has not been explored before. The results indicate that the exceptional-point physics can be used to control photon-photon interaction, and therefore the exceptional-point physics may prove useful in the quest to create many-body quantum photon states in waveguide QED systems.

This research is supported by AFOSR-MURI programs, Grant No. FA9550-12-1-0488 and  FA9550-17-1-0002.

\newpage

\section{Supplementary Information}
When the input state consists of $N$ photons with frequencies $k_1, k_2, \cdots, k_N$, the wavefunction of output $N$-photon bound state is
\begin{eqnarray}\label{B}
B\left(x_1, \cdots, x_N\right) =\int \frac{dp_1\cdots dp_N}{\left(\sqrt{2\pi}\right)^N}  S_{p_1\cdots p_Nk_1\cdots k_N}^C \frac{1}{\sqrt{N!}}\sum_{Q}e^{ip_{1}x_{Q(1)}}\cdots e^{ip_{N}x_{Q(N)}}\,,
\end{eqnarray}
where $Q$ denotes all the permutations on indices $\{1,\cdots,N\}$ and $S_{p_1\cdots p_Nk_1\cdots k_N}^C$ is the connected part of $N$-photon S matrix \cite{nphoton}. 
Here, we focus on the decay behavior of $N$-photon bound state as a function of photons' separations. For this purpose, all we have to do is to identify the pole structures 
of $p_1,\cdots, p_N$ in $S_{p_1\cdots p_N k_1\cdots k_N}^C$, which can be written down directly in a diagrammatical approach as proposed in Ref. \cite{diagram}. 

\begin{figure}[h]
\includegraphics[width=0.9\textwidth]{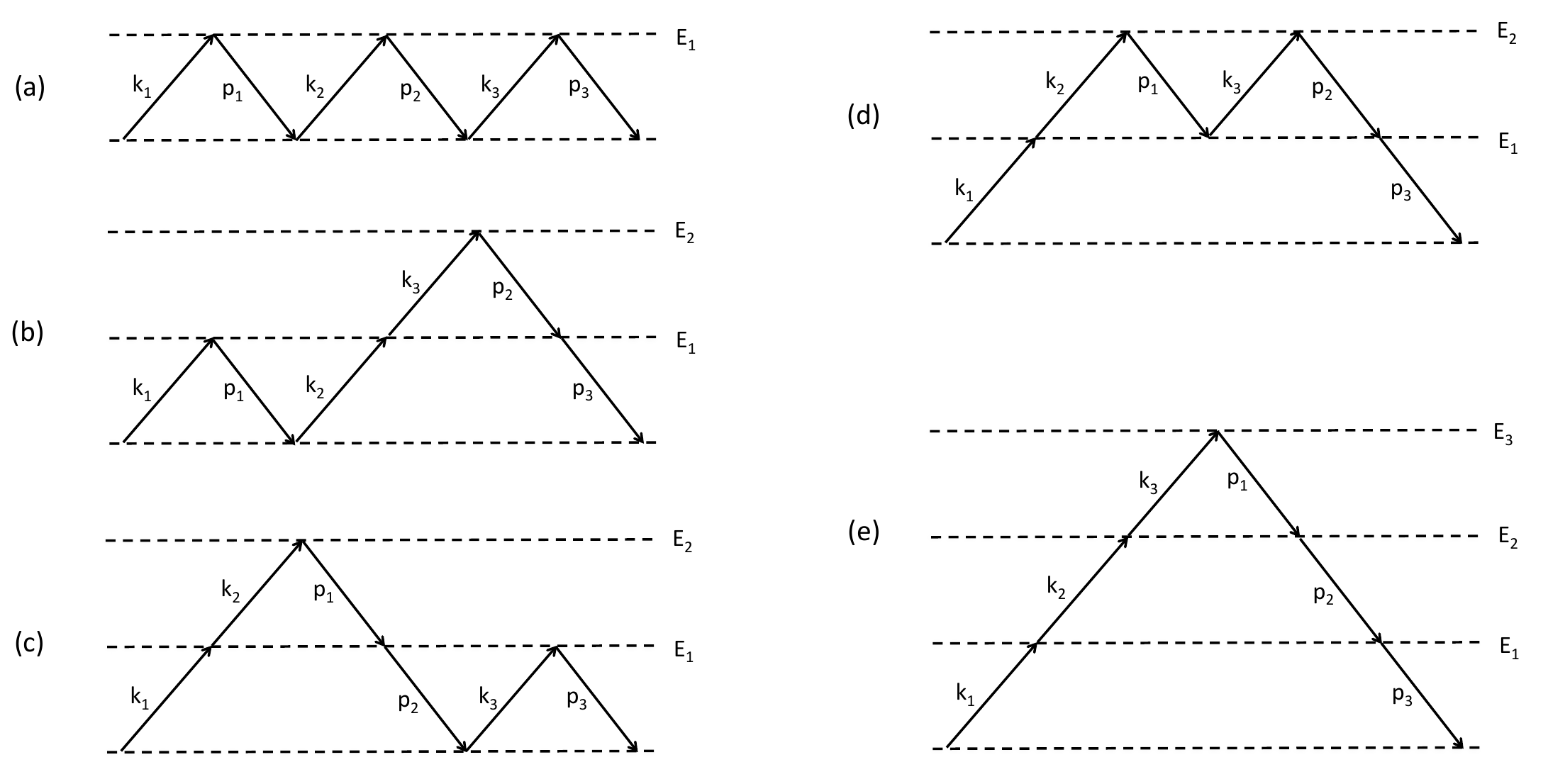}
\caption{The five types of terms of three-photon S matrix in diagrams. (a) $\langle aa^{\dag}aa^{\dag}aa^{\dag}\rangle$; (b) $\langle aaa^{\dag}a^{\dag}aa^{\dag}\rangle$; 
(c) $\langle aa^{\dag}aaa^{\dag}a^{\dag}\rangle$; (d) $\langle aaa^{\dag}aa^{\dag}a^{\dag}\rangle$; (e) $\langle aaaa^{\dag}a^{\dag}a^{\dag}\rangle$.}\label{sm_fig}
\end{figure}

For illustration,
we take the three-photon case as an example. Our discussion can be generalized straightforwardly to the $N$-photon case.  
For three photons, $S^C_{p_1p_2p_3k_1k_2 k_3}$ is the sum of five diagrams as listed in Fig. \ref{sm_fig} up to permutations of photon frequencies.
Following the rules in \cite{diagram}, for each diagram in Fig. \ref{sm_fig}, we define variables 
\begin{eqnarray}
P_1\equiv p_1\,,\,\,\,\, P_2\equiv p_1+p_2\,,\,\,\,\,P_3\equiv p_1+p_2+p_3\,,\,\,\,\,K_1\equiv k_1\,,\,\,\,\,K_2\equiv k_1+k_2\,,\,\,\,\,K_3\equiv k_1+k_2+k_3\,,\nonumber
\end{eqnarray}
and write down the pole structure as follows :
\begin{enumerate}[label=(\alph*)]
\item $\frac{1}{K_1-E_1}\frac{{\cal{P}}}{K_1-P_1}\frac{1}{K_2-P_1-E_1}\frac{{\cal{P}}}{K_2-P_2}\frac{1}{K_3-P_2-E_1}\delta(K_3-P_3)$, 
\item $\frac{1}{K_1-E_1}\frac{{\cal{P}}}{K_1-P_1}\frac{1}{K_2-P_1-E_1}\frac{1}{K_3-P_1-E_2}\frac{1}{K_3-P_2-E_1}\delta(K_3-P_3)$,
\item $\frac{1}{K_1-E_1}\frac{1}{K_2-E_2}\frac{1}{K_2-P_1-E_1}\frac{{\cal{P}}}{K_2-P_2}\frac{1}{K_3-P_2-E_1}\delta(K_3-P_3)$,
\item $\frac{1}{K_1-E_1}\frac{1}{K_2-E_2}\frac{1}{K_2-P_1-E_1}\frac{1}{K_3-P_1-E_2}\frac{1}{K_3-P_2-E_1}\delta(K_3-P_3)$,
\item $\frac{1}{K_1-E_1}\frac{1}{K_2-E_2}\frac{1}{K_3-E_3}\frac{1}{K_3-P_1-E_2}\frac{1}{K_3-P_2-E_1}\delta(K_3-P_3)$.
\end{enumerate}
Based on the poles listed above, we can evaluate (\ref{B}) by contour integrals with respect to new variables $P_1$, $P_2$ and $P_3$, which gives the decay forms:
\begin{enumerate}[label=(\alph*)]
\item	$e^{-iE_1\left(x_{Q(1)}-x_{Q(2)}\right)}e^{iE_1\left(x_{Q(3)}-x_{Q(2)}\right)}\theta(x_{Q(1)}-x_{Q(2)})\theta(x_{Q(2)}-x_{Q(3)})$,  
\item $\left[e^{-iE_1\left(x_{Q(1)}-x_{Q(2)}\right)}+ e^{-iE_2\left(x_{Q(1)}-x_{Q(2)}\right)}\right]e^{iE_1\left(x_{Q(3)}-x_{Q(2)}\right)}\theta(x_{Q(1)}-x_{Q(2)})\theta(x_{Q(2)}-x_{Q(3)})$,  
\item $e^{-iE_1\left(x_{Q(1)}-x_{Q(2)}\right)}e^{iE_1\left(x_{Q(3)}-x_{Q(2)}\right)}\theta(x_{Q(1)}-x_{Q(2)})\theta(x_{Q(2)}-x_{Q(3)})$,  
\item  $\left[e^{-iE_1\left(x_{Q(1)}-x_{Q(2)}\right)}+ e^{-iE_2\left(x_{Q(1)}-x_{Q(2)}\right)}\right]e^{iE_1\left(x_{Q(3)}-x_{Q(2)}\right)}\theta(x_{Q(1)}-x_{Q(2)})\theta(x_{Q(2)}-x_{Q(3)})$,  
\item $e^{-iE_2\left(x_{Q(1)}-x_{Q(2)}\right)}e^{iE_1\left(x_{Q(3)}-x_{Q(2)}\right)}\theta(x_{Q(1)}-x_{Q(2)})\theta(x_{Q(2)}-x_{Q(3)})$.  
\end{enumerate}
Summing them together, the wavefunction of the three-photon bound state $B(x_1, x_2, x_3)$ has the decay form of 
\begin{equation}
B(x_1,x_2,x_3)\sim \sum_{Q}\left[A\,e^{-iE_1\left(x_{Q(1)}-x_{Q(2)}\right)}+B\, e^{-iE_2\left(x_{Q(1)}-x_{Q(2)}\right)}\right]e^{-iE_1\left(x_{Q(2)}-x_{Q(3)}\right)}\theta(x_{Q(1)}-x_{Q(2)})\theta(x_{Q(2)}-x_{Q(3)})\,.
\end{equation} 

In general, for $N$ photons, the wavefunction of the $N$-photon bound state always has the pairwise decay form as 
\begin{eqnarray}\label{pairwise}
B\left(x_1, \cdots, x_N\right) \propto \sum_Q\,\prod_{i=1}^{N-1} D_i\left(x_{Q(i)}-x_{Q(i+1)}\right)\theta\left(x_{Q(i)}-x_{Q(i+1)}\right)\,,
\end{eqnarray}
where each $D_i\left(x_{Q(i)}-x_{Q(i+1)}\right)$ is the linear combination of exponential decays controlled by different excitation poles. The proof is similar to that in previous three-photon case. 
For each diagram of the connected $N$-photon S matrix, we assign the up arrows from left to right with labels $k_1,\cdots, k_N$ and  the down arrows from left to right with labels  
$p_1,\cdots, p_N$. The connected $N$-photon S matrix is the direct product of terms,  each containing one of the poles of $P_1\equiv p_1, P_2\equiv p_1+p_2,\cdots, P_{N-1}\equiv p_1+\cdots+p_{N-1}$. To evaluate (\ref{B}), 
we first integrate out $p_N$ to remove the $\delta$-function and the exponential term $e^{ip_{1}x_{Q(1)}}\cdots e^{ip_{N}x_{Q(N)}}$ in (\ref{B}) becomes the form of 
$e^{iP_{1}(x_{Q(1)}-x_{Q(2)})}e^{iP_{2}(x_{Q(2)}-x_{Q(3)})}\cdots e^{iP_{N-1}(x_{Q(N-1)}-x_{Q(N)})}$. As a result, the integral (\ref{B}) is decomposed to $N-1$ independent integrals with 
respect to variables $P_1,\cdots, P_{N-1}$, which results in the pairwise form as shown in (\ref{pairwise}).

For the general form of (\ref{pairwise}), because of the existence of the diagram like Fig.\ref{sm_fig} (a) that only contains the single excitation poles, 
$D_i\left(x_{Q(i)}-x_{Q(i+1)}\right)$ must at least contain the term $e^{-i E_1\left(x_{Q(i)}-x_{Q(i+1)}\right)}$. In typical waveguide QED systems, $\text{Im}E_i < \text{Im} E_1 < 0$ for $i > 1$,  
the single excitation poles $E_1$ dominates the decay (the slowest decay mode). As a result, the slowest decay part in (\ref{pairwise}) has the form:
\begin{eqnarray}\label{slow}
B^{\text{(slowest)}}\left(x_1, \cdots, x_N\right) \propto \sum_Q\,\prod_{i=1}^{N-1} e^{-i E_1\left(x_{Q(i)}-x_{Q(i+1)}\right)}\theta\left(x_{Q(i)}-x_{Q(i+1)}\right)\,.
\end{eqnarray}

\begin{figure}[h]
\includegraphics[width=0.8\textwidth]{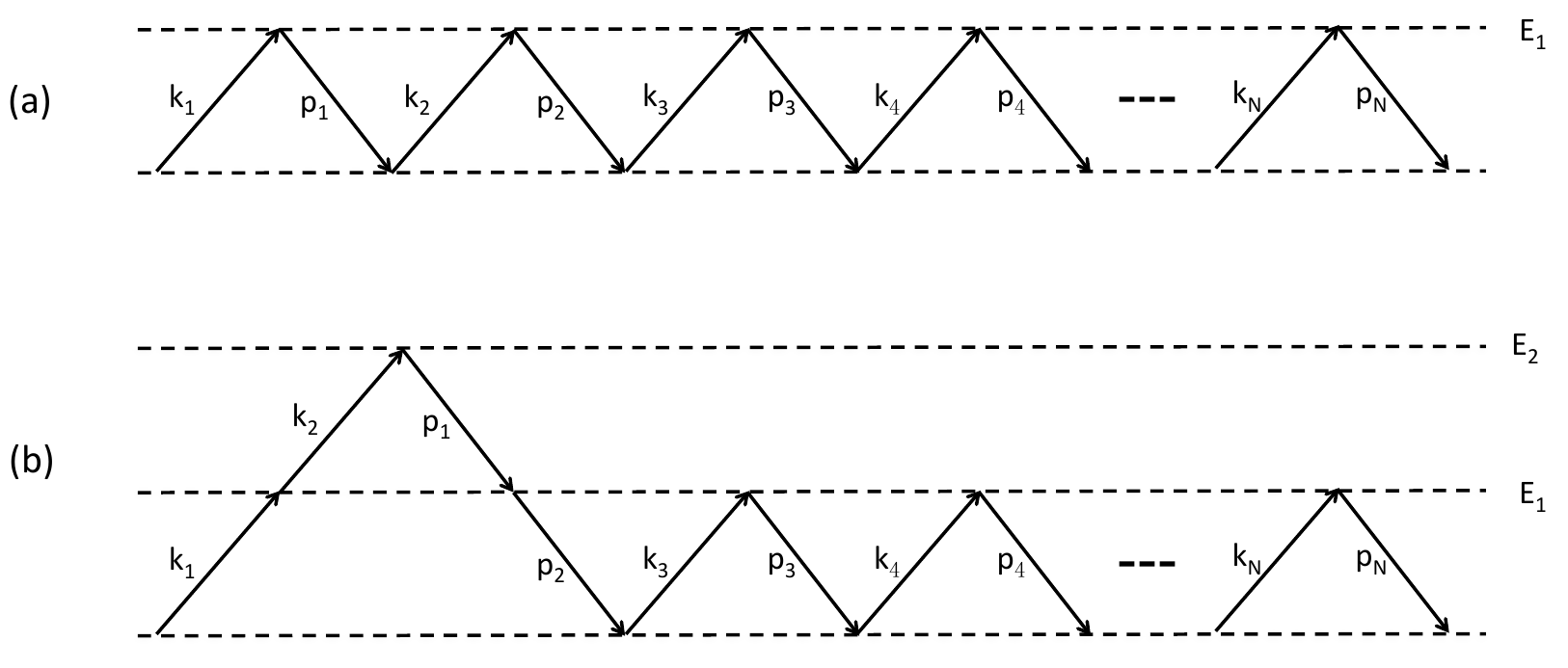}
\caption{The only two types of diagrams that contain only single excitations pole of $P_1,P_2,\cdots, P_{N-1}$. (a) $\langle aa^{\dag} \cdots  aa^{\dag} aa^{\dag}aa^{\dag}aa^{\dag}\rangle$; 
(b) $\langle aa^{\dag} \cdots  aa^{\dag} aa^{\dag}aaa^{\dag}a^{\dag}\rangle$.} \label{sm_fig3}
\end{figure}

Furthermore, we can calculate the exact form of the slowest decay (\ref{slow}) explicitly. It can be proved that there are only two types of diagrams, as listed in Fig.\ref{sm_fig3}, 
in which all the poles of $P_1, P_2,\cdots, P_{N-1}$ are single excitation poles.  Summing up terms corresponding to these two diagrams lead to (\ref{slow}). 
The exact form of the connected part of $N$-photon S matrix contributed by the two diagrams is
\begin{eqnarray} 
S^{C\,\text{(slowest)}}_{p_1\cdots p_Nk_1\cdots k_N}= \frac{\delta(P_N-K_N)}{(2\pi i)^{N-2}} \sum_{Q, R}{\cal{G}}\left(P_{Q(1)},K_{R(1)},K_{R(2)}\right)
\prod_{j=2}^{N-1}\frac{{\cal{P}}}{P_{Q(j)} - K_{R(j)}}G\left(K_{R(j+1)}- P_{Q(j)} \right)\,,
\end{eqnarray}
where $Q, R$ are the permutations on indices $\{1,\cdots, N\}$. $K_{R(i)}\equiv \sum_{l=1}^i k_{R(l)}$ and $P_{Q(i)}\equiv \sum_{l=1}^i p_{Q(l)}$ for $i=1,\cdots, N$.
$G(k)$ is related to the single photon S matrix as $S_{pk} = \left[1 + G(k)\right]\delta(p-k)$ and
${\cal{G}}\left(P_1,K_1,K_2\right)$ is related to the connected two-photon S matrix as
\begin{eqnarray}
S_{p_1p_2k_1,k_2}^C =\left[{\cal{G}}\left(p_1,k_1, k_1+k_2\right)+{\cal{G}}\left(p_2, k_1,k_1+k_2\right)+
{\cal{G}}\left(p_1,k_2, k_1+k_2\right)+{\cal{G}}\left(p_2,k_2, k_1+k_2\right)\right]\delta(p_1+p_2-k_1-k_2)\,.\nonumber
\end{eqnarray}
For the effective Hamiltonian $H_{\text{eff}}=H_{\text{loc}}-i\frac{\kappa}{2}a^{\dag}a$,  $H_{\text{eff}}|\lambda\rangle_n = E_{n\lambda}|\lambda\rangle_n$, $N_{\text{loc}}|\lambda\rangle_n = n|\lambda\rangle_n$, 
\begin{eqnarray}
G(k) &=& -i \kappa \sum_{\lambda}\frac{\langle 0|a|\lambda\rangle_1 {}_1\langle \bar{\lambda} |a^{\dag}|0\rangle }{k-E_{1\lambda}}\,,\nonumber\\
{\cal{G}}\left(P_1,K_1,K_2\right)&=&\frac{\kappa^2}{2\pi i}\sum_{\mu\nu}\frac{\langle 0|a|\nu\rangle_1}{K_2-P_1-E_{1\nu}}
\left[{}_1\langle\bar{\nu} |a^{\dag}|0\rangle\frac{\cal{P}}{K_1-P_1}\langle 0|a|\mu\rangle_1-\sum_{\lambda}
\frac{{}_1\langle \bar{\nu}|a|\lambda\rangle_2 {}_2\langle \bar{\lambda} |a^{\dag}|\mu\rangle_1}{K_2-E_{2\lambda}} \right]
\frac{{}_1\langle\bar{\mu} |a^{\dag}|0\rangle}{K_1-E_{1\mu}}\,.\nonumber
\end{eqnarray}
The wavefunction of the slowest decay part of the bound state can be evaluated as
\begin{eqnarray}
B^{\text{(slowest)}}\left(x_1, \cdots, x_N\right)&=&\int \frac{dp_1\cdots dp_N}{\left(\sqrt{2\pi}\right)^N}  S_{p_1\cdots p_Nk_1\cdots k_N}^{C\,\text{(slowest)}} \frac{1}{\sqrt{N!}}\sum_{Q}e^{ip_{1}x_{Q(1)}}\cdots e^{ip_{N}x_{Q(N)}}\nonumber\\
&=&\frac{\sqrt{N!}}{\left(\sqrt{2\pi}\right)^N}\sum_{Q,R}e^{iK_Nx_{Q(N)}}\left(\int dP_1\, {\cal{G}}\left(P_{1},K_{R(1)},K_{R(2)}\right) e^{iP_1(x_{Q(1)}-x_{Q(2)})} \right)\nonumber\\
&&\,\,\,\,\,\,\,\,\,\,\,\,\,\,\,\,\,\,\,\,\,\,\,\,\,\,\,\,\,\,\,\,\,\,\,\,\,\,
\times\prod_{j=2}^{N-1}\int \frac{dP_j}{2\pi i}\frac{{\cal{P}}}{P_{j} - K_{R(j)}}G\left(K_{R(j+1)}- P_{j} \right)e^{iP_j(x_{Q(j)}-x_{Q(j+1)})}\,.\label{main}
\end{eqnarray}

When there are two single excitation poles $E_{1,+}$ and $E_{1,-}$,  $G(k)$ and ${{\cal{G}}\left(P_{1},K_{1},K_{1}\right)}$ have the form of 
\begin{eqnarray}
G(k)\equiv\frac{{\cal{A}}(k)}{\left(k-E_{1,+}\right)\left(k-E_{1,-}\right)}\,,\,\,\,\,\,\,{{\cal{G}}\left(P_{1},K_{1},K_{2}\right)}\equiv\frac{1}{2\pi i }\frac{{\cal{B}}(P_{1},K_{1},K_{2})}
{\left(K_2-P_1-E_{1,+}\right)\left(K_2-P_1-E_{1,-}\right)}\,,
\end{eqnarray}
where ${\cal{A}}(k)$  and ${\cal{B}}(P_{1},K_{1},K_{2})$ are the analytic function of variables $k$ and $P_1$, respectively. We can evaluate
\begin{eqnarray}
&&\int dP_1\, {\cal{G}}\left(P_{1},K_{R(1)},K_{R(2)}\right) e^{iP_1(x_{Q(1)}-x_{Q(2)})}\nonumber\\
&=&\int \frac{dP_1}{2\pi i }\frac{{\cal{B}}(P_{1},K_{R(1)},K_{R(2)})}{\left(K_{R(2)}-P_1-E_{1,+}\right)\left(K_{R(2)}-P_1-E_{1,-}\right)}e^{iP_1(x_{Q(1)}-x_{Q(2)})}\nonumber\\
&=&\left[{\cal{B}}(K_{R(2)}-E_{1,+}, K_{R(1)},K_{R(2)})\,e^{-iE_{1,+}(x_{Q(1)}-x_{Q(2)})}-{\cal{B}}(K_{R(2)}-E_{1,-},K_{R(1)},K_{R(2)})\,e^{-iE_{1,-}(x_{Q(1)}-x_{Q(2)})}
\right]\nonumber\\
&& \,\,\times\frac{e^{iK_{R(2)}(x_{Q(1)}-x_{Q(2)})}}{E_{1,+}-E_{1,-}}\theta(x_{Q(1)}-x_{Q(2)})\nonumber\\
&\equiv&F_{k_{R(1)}, k_{R(2)}}\left(x_{Q(1)}-x_{Q(2)}\right)e^{iK_{R(2)}(x_{Q(1)}-x_{Q(2)})}\theta(x_{Q(1)}-x_{Q(2)})\,,\label{two-decay}
\end{eqnarray}
and
\begin{eqnarray}
&&\int \frac{dP_j}{2\pi i} \frac{{\cal{P}}}{P_{j} - K_{R(j)}}G\left(K_{R(j+1)}- P_{j} \right)e^{iP_j(x_{Q(j)}-x_{Q(j+1)})}\nonumber\\
&=&\int \frac{dP_j}{2\pi i} \frac{{\cal{P}}}{P_{j} - K_{R(j)}}\frac{{\cal{A}}(K_{R(j+1)}- P_{j} )}{\left(P_j-K_{R(j+1)}+E_{1,+}\right)\left(P_j-K_{R(j+1)}+E_{1,-}\right)}e^{iP_j(x_{Q(j)}-x_{Q(j+1)})}\nonumber\\
&=&\left[\frac{{\cal{A}}(E_{1,+})\,e^{-iE_{1,+}(x_{Q(j)}-x_{Q(j+1)})}}{k_{R(j+1)}-E_{1,+}}
-\frac{{\cal{A}}(E_{1,-})\,e^{-iE_{1,-}(x_{Q(j)}-x_{Q(j+1)})}}{k_{R(j+1)}-E_{1,-}}
\right]\frac{e^{iK_{R(j+1)}(x_{Q(j)}-x_{Q(j+1)})}}{E_{1,+}-E_{1,-}}\theta(x_{Q(j)}-x_{Q(j+1)})\nonumber\\
&\equiv& F_{k_{R(j+1)}}\left(x_{Q(j)}-x_{Q(j+1)}\right)e^{iK_{R(j+1)}(x_{Q(j)}-x_{Q(j+1)})}\theta(x_{Q(j)}-x_{Q(j+1)})\,.\label{single-decay}
\end{eqnarray} 
As a result,
\begin{eqnarray}
B^{\text{slowest}}(x_1, \cdots, x_N)
&=&\frac{\sqrt{N!}}{\left(\sqrt{2\pi}\right)^N}\sum_{Q,R}F_{k_{R(1)}, k_{R(2)}}\left(x_{Q(1)}-x_{Q(2)}\right)e^{i(k_{R(1)}+k_{R(2)})x_{Q(1)}}\theta(x_{Q(1)}-x_{Q(2)})\nonumber\\
&&\,\,\,\,\,\,\times \prod_{j=2}^{N-1}F_{k_{R(j+1)}}\left(x_{Q(j)}-x_{Q(j+1)}\right)
e^{ik_{R(j+1)}x_{Q(j)}}\theta(x_{Q(j)}-x_{Q(j+1)})\,.\label{result1}
\end{eqnarray}

For the special case of Jaynes-Cummings model, we have
\begin{eqnarray}
E_{1,\pm} =  \frac{\left(\omega-i\frac{\kappa}{2}\right)+\Omega}{2}\pm\sqrt{\left(\frac{\omega-i\frac{\kappa}{2}-\Omega}{2}\right)^2+g^2}\,,\,\,\,\,\,\,\,\,\,\,\,\,\,\,\,\,
{\cal{A}}(k)= -i\,\kappa\left(k-\Omega\right)\,.\nonumber
\end{eqnarray}
Consider the resonsant scattering $k_1=k_2=\cdots=k_N=\omega=\Omega$, as shown in the main context, (\ref{two-decay}) has the decay form of $f(x_{Q(1)}-x_{Q(2)})$ while
(\ref{single-decay}) has the decay form of $g(x_{Q(j)}-x_{Q(j+1)})$ with $g(\tau)$ defined as
\begin{equation}\label{gtau}
g(\tau)\equiv \begin{cases}
\frac{\sin\sqrt{g^2-\left(\frac{\kappa}{4}\right)^2}|\tau| }{\sqrt{(g)^2-\left(\frac{\kappa}{4}\right)^2}}e^{-\frac{\kappa}{4}|\tau|}   & \kappa<4g  \\
|\tau| e^{-g|\tau|} \ &\kappa=4g \\
\frac{\sinh\sqrt{\left(\frac{\kappa}{4}-g^2\right)^2}|\tau| }{\sqrt{\left(\frac{\kappa}{4}\right)^2-g^2}}e^{-\frac{\kappa}{4}|\tau|} & \kappa>4g  \\
\end{cases}\,.
\end{equation}
Therefore, the $N$-photon bound state scattered from Jaynes-Cummings system in the resonant scattering case has the decay form of
 \begin{eqnarray}
B(x_1, \cdots, x_N) \propto \sum_{Q}f(x_{Q(j)}-x_{Q(j+1)})\theta(x_{Q(1)}-x_{Q(2)}) \prod_{j=2}^{N-1}g(x_{Q(j)}-x_{Q(j+1)})\theta(x_{Q(j)}-x_{Q(j+1)})
\end{eqnarray}

\end{document}